\documentstyle[12pt]{article}
\def\lapproxeq{\lower .7ex\hbox{$\;\stackrel{\textstyle
<}{\sim}\;$}}
\def\gapproxeq{\lower .7ex\hbox{$\;\stackrel{\textstyle
>}{\sim}\;$}}

\begin{document}

\begin{center}
{\bf ON MEASURING THE $\gamma \gamma$ WIDTH OF THE INTERMEDIATE-
\\ MASS HIGGS AT A PHOTON LINEAR COLLIDER}
\end{center}

\vspace*{1cm}
\begin{center}
VALERY A.\ KHOZE\footnote{Talk at the Photon' 95 Conference,
Sheffield, April 8-15, 1995.
Based on work together with D.\ Borden,
W.\ J.\ Stirling and J.\ Ohnemus \cite{BSO}} \\
Department of Physics, University of Durham, \\ Durham, DH1 3LE,
U.K.
\end{center}

\vspace*{1cm}

\begin{abstract}
The identification of the intermediate-mass Higgs process $\gamma
\gamma \rightarrow H \rightarrow b\overline{b}$ will be one of
the most important goals of a future photon linear collider.
Backgrounds from the continuum $\gamma \gamma \rightarrow
c\overline{c}, b\overline{b}$ leading-order processes can be
suppressed by using polarized photon beams in the $J_z = 0$
initial-state.  The radiative processes $\gamma \gamma
\rightarrow c\overline{c} g, b\overline{b} g$ can mimic the
two-jet topology of the Higgs signal and provide the dominant
background in the $J_z = 0$ channel.  Particularly problematic is
the $\gamma \gamma \rightarrow c\overline{c} g \rightarrow 2$
jets process.  The effects of imposing additional cuts are
investigated and it is shown that the radiative background could
be reduced to a manageable level.
\end{abstract}

\vspace*{2cm}

\noindent  {\Large \bf 1.  Introduction}

\medskip
The rapid advance of laser technology makes possible the
collision of high- \linebreak brightness, high-energy photon
beams at future linear colliders [2-3] through Compton
backscattering.  One particularly intriguing use of such a photon
linear collider is to measure the two-photon decay with a Higgs
boson, $\Gamma (H \rightarrow \gamma\gamma)$ [4-6].  In a photon
linear collider $\Gamma (H \rightarrow \gamma\gamma)$ is deduced
by measuring the Higgs production cross section in the reaction
$\gamma\gamma \rightarrow H \rightarrow X$ where $X$ is the
detected final state.  The number of events is proportional to
the product $\Gamma (H \rightarrow \gamma\gamma) \: B(H
\rightarrow X)$ where $B(H \rightarrow X)$ is the branching ratio
of the Higgs boson into the final state $X$.

For a Higgs boson in the intermediate-mass region, 50 GeV
\lapproxeq 150 GeV, the dominant decay mode is to
$b\overline{b}$.  Measurement of the two-photon width in this
mass region requires suppressing the continuum $\gamma\gamma
\rightarrow b\overline{b}, c\overline{c}$ background beneath the
resonant $\gamma\gamma \rightarrow H \rightarrow b\overline{b}$
signal, assuming light quarks can be distinguished from heavy
quarks by vertexing.  The background can be greatly suppressed by
using polarized photon beams.  The Higgs signal is produced by
photons in a $J_z = 0$ initial state, whereas the backgrounds are
primarily produced by photons in the $J_z = \pm 2$ initial state,
the $J_z = 0$ Born cross section being suppressed for large
angles by a factor of $m_q^2/s$ \cite{GH,KAI}.

The radiative processes can have an important impact (\lq\lq
radiation damage") since the presence of the additional gluon(s)
in principle removes the suppression of the $J_z = 0$ background
channel \cite{BSO,JT}.  The radiative process $\gamma\gamma
\rightarrow q\overline{q} g$ with $q = b,c$ can mimic the
dominant two-jet topology of the Higgs signal in two important
cases:  (i) if two of the three partons are collinear, for
example a fast quark recoiling against a collinear quark and
gluon, or (ii) if one of the partons is either quite soft or is
directed down the beampipe and is therefore not tagged as a
distinct jet.  A particularly interesting example of the latter
is when one of the incoming photons splits into a quark and an
antiquark, one of which carries most of the photon's momentum and
Compton scatters off the other photon, $q(\overline{q}) \gamma
\rightarrow q(\overline{q}) g$.  Two jets are then identified in
the detector, with the third jet remaining undetected.  Here we
discuss the impact of the radiative $q\overline{q} g$ process on
the study of an intermediate-mass Higgs boson at a photon linear
collider.

\bigskip
\noindent {\Large \bf 2.  Non-radiative processes}

\medskip
For Higgs boson in the intermediate-mass region, the beam energy
spread is much greater than the total width, and so the number of
$H \rightarrow b\overline{b}$ events expected is
\begin{equation}
N_{H \rightarrow b\overline{b}} = \left . \frac{d
L_{\gamma\gamma}^{J_z = 0}}{d W_{\gamma\gamma}} \right |_{M_H} \:
\frac{8 \pi^2 \Gamma (H \rightarrow \gamma\gamma) B (H
\rightarrow b\overline{b})}{M_H^2}
\label{a1}
\end{equation}
\noindent where $W_{\gamma\gamma} = \sqrt s$ is the two-photon
invariant mass.

In the Born approximation the background $(\gamma\gamma
\rightarrow b\overline{b}, c\overline{c})$ cross section is given
by
\begin{eqnarray}
\frac{d \sigma (\gamma\gamma \rightarrow q\overline{q})}{d \cos
\theta} & = & \frac{12 \pi \alpha^2 Q_q^4}{s} \: \frac{\beta}{(1
- \beta^2 \cos^2 \theta)^2} \label{a2} \\
& & \times \left \{ \begin{array}{ll}
1 - \beta^4 & {\rm for} \: J_z = 0 \\
\beta^2 (1 - \cos^2 \theta)(2 - \beta^2 + \beta^2 \cos^2 \theta)
& {\rm for} \: J_z = \pm 2 \end{array} \right . \nonumber
\end{eqnarray}
\noindent where $\beta \equiv \sqrt{1 - 4 m_q^2/s}$ and $m_q$ and
$Q_q$ are the mass and electric charge of the quark respectively.

Note the strong $\cos \theta$ dependence of the cross section and
that the $J_z = 0$ cross section vanishes, for $|\cos \theta| <
1$, in the high-energy limit.  This cross section can therefore
be significantly reduced by using polarized beams and cutting on
$\cos \theta$.  Note that in comparing signal ($S$) to background
($B$) cross sections it is convenient to normalize the signal
cross sections as if $(dL/dW)_S = (L)_B$/(10 GeV).  This is
equivalent to assuming that the experimental resolution on
reconstructing the Higgs mass is 10 GeV, see \cite{BBC}.

Fig.\ 1 shows the Born cross sections for $b\overline{b}$ and
$c\overline{c}$ production and illustrates the very large
suppression that is possible with polarized photons in the $J_z =
0$ state.

Before discussing the radiative background we comment on the
origin of the large-angle suppression of $q\overline{q}$
production in the $J_z = 0$ channel as $m_q^2/s \rightarrow 0$.
Consider the symmetry properties of the Born amplitude in the
$\beta \rightarrow 1$ limit.  Because of helicity conservation at
the photon vertices, only amplitudes with opposite helicities for
the quark and antiquark survive.  However, the combined impact of
$C-, P-$ and $T-$invariance, photon Bose statistics, and
unitarity, can be shown to lead to a vanishing amplitude in this
limit {\it at lowest order in perturbation theory} \cite{BSO}.

There are some instructive lessons from studying these symmetry
properties:
\begin{enumerate}
\item For the $J_z = 0$ case the interferences between the Born
and higher-order non-radiative diagrams are additionally
suppressed by a factor of $m_q/\sqrt s$.
\item For $b\overline{b}$ production at $W_{\gamma\gamma}
\sim$ 100 GeV the suppression factor $\frac{m_b}{\sqrt s} \sim
\frac{\alpha_S}{\pi}$ and the virtual gluon ${\cal O}
(\alpha_S^2)$ corrections should be taken into account in
computation of the two-jet background in the $J_z = 0$
channel\footnote{In the $b\overline{b}$ case the lowest-order
results of Refs.\ \cite{BSO,BBC} could be considered only as the
qualitative guide allowing one to estimate the scale of the
background effects (see also \cite{JT}).  In the $c\overline{c}$
case which is our main concern the tree-level calculations are
quite appropriate.}.
\item For the special case of scattering at $90^o$, the
$\frac{m_q}{\sqrt s}$-suppression of all $J_z = 0$ non-radiative
amplitudes (i.e.\ not just at leading order) follows simply from
rotational invariance about the fermion direction and photon Bose
statistics.  For this particular angular configuration the $T$-
invariance argument is redundant.
\end{enumerate}

\bigskip
\noindent {\Large \bf 3.  Radiative background}

\medskip
While at sufficiently high energies the lowest order
$q\overline{q}$ large-angle cross sections are ${\cal O}
(\alpha^2/s)$ and ${\cal O} (\alpha^2 m_q^2/s^2)$ for $J_z = \pm
2$ and 0, respectively, the $q\overline{q}g$ cross sections are
${\cal O} (\alpha^2 \alpha_S/s)$ in both cases, i.e.\ the
$\gamma\gamma \rightarrow q\overline{q}g$ cross section is in
principle {\it not} suppressed in the $J_z = 0$ channel.
Furthermore, there are regions of phase space where the three-
parton final state may be tagged as a two-jet event.  In the case
of $b\overline{b}g$ and $c\overline{c}g$, the event may have a
vertex structure similar to the non-radiative case, in which case
this process could easily be misidentified as a $b\overline{b}$
final state.  In contrast, the $J_z = \pm 2$ cross section for
$\gamma\gamma \rightarrow q\overline{q}g$ is simply an ${\cal O}
(\alpha_S)$ correction to the much larger $J_z = \pm 2 \:
\gamma\gamma \rightarrow q\overline{q}$ cross section and will
not be considered further here.

In Ref.\ \cite{BSO} we discussed the various radiative
contributions which could be tagged as two-jet events.  The
analytic formulae were presented in the limit of vanishing quark
masses and the results for the realistic case of massive quarks
were calculated numerically.  Recall that in the total cross
section the infrared singularity is cancelled by one-loop virtual
gluon correction to the Born cross section.

The largest background is from the $\gamma\gamma \rightarrow
c\overline{c}g$ process which results in an order of magnitude
increase of the $J_z = 0$ two-jet production cross section at
$\sqrt s \sim$ 100 GeV.  This cross section is, in principle, in
excess of the Higgs signal.  An important contribution
corresponds to one of the photons splitting into a quark and an
antiquark, one of which undergoes a Compton scattering with the
other photon to produce an energetic quark and gluon.  The extent
to which these two jets are back-to-back in the $\gamma\gamma$
center-of-mass frame (and therefore constitute a background to $H
\rightarrow q\overline{q}$) depends on how the momentum is
apportioned between the active and spectator quark in the $\gamma
\rightarrow q\overline{q}$ splitting.  It is worth noting that
there is no $J_z = 0$ suppression in this case.

To estimate the size of the virtual Compton scattering
contribution, one can use the leading pole approximation
\cite{BFK}, i.e.
\begin{eqnarray}
d \sigma (\gamma\gamma \rightarrow q\overline{q}g) & \simeq & d
{\cal W} (\gamma \rightarrow q\overline{q}) \: d \sigma (q \gamma
\rightarrow q g)|_{p^* = k_1 - \overline{p}}, \label{a3} \\
\nonumber \\
d {\cal W} (\gamma \rightarrow q\overline{q}) & = & \frac{\alpha
Q_q^2}{4 \pi^2} \: \left [ \frac{\overline{x}^2 + (1 -
\overline{x})^2}{k_1 . \overline{p}} + \frac{(1 - \overline{x})
m_q^2}{(k_1 . \overline{p})^2} \right ] \: \frac{d^3
\overline{p}}{\overline{p}_0}, \label{a4}
\end{eqnarray}
\noindent where $\overline{x} = 2 \overline{p}_0/\sqrt s$ is the
energy fraction of the quark which does not participate in the
hard scattering.  For this process to give a two-jet background,
most of the $\gamma\gamma$ scattering energy $\sqrt s$ should be
deposited in the detector, thus $0 < \overline{x} < \epsilon$
where $\epsilon$ is a small parameter that will be directly
related to the allowed acollinearity of the two jets in the
detector.  Recall that discriminating two-jet topology on an
event-by-event basis requires specifying a jet-finding algorithm.

It is convenient to use a clustering formalism exemplified by the
JADE algorithm \cite{JADE}.  If we use this algorithm $\epsilon
\sim y_{cut}$ and the pure $q\overline{q}$ events could be
efficiently tagged with a $y_{cut}$ of 0.02--0.03.

The transverse momentum integration of the spectator quark gives
rise to a large logarithm, $\sim \ln (\Delta s/m_q^2)$, where
$\Delta s$ is some fraction of $s$, and so the overall size of
this contribution is roughly
\begin{eqnarray}
\sigma (\gamma\gamma \rightarrow q\overline{q}g \rightarrow 2 \:
\hbox{jets})_{\rm comp} & \simeq & \frac{\alpha Q_q^2}{2 \pi}
\: {\cal O} (\epsilon) \: \ln \left (\frac{\Delta s}{m_q^2}
\right ) \: \sigma (q \gamma \rightarrow qg) \simeq \nonumber \\
\label{a5} \\
& \simeq & \frac{\alpha^2}{s} \alpha_S {\cal O} (y_{cut})
\: \ln \left (\frac{\Delta s}{m_q^2} \right ). \nonumber
\end{eqnarray}

Note that the requirement that most of the collision energy
should be deposited at large angles in the detector in the case
of monochromatic photon beams provides a very strong suppression
of other \lq resolved photon' contributions, such as $\gamma
\rightarrow gX$ followed by $g \gamma \rightarrow q\overline{q}$
\cite{VIT}.

Recall that for $\sqrt s \simeq$ 100 GeV, $\frac{mb}{\sqrt s}
\sim \frac{\alpha_S}{\pi} \sim y_{cut}$ and the Born contribution
and the virtual and gluon radiative corrections to the two-jet
cross section are all of the same order.

\newpage
\noindent {\Large \bf 4.  Experimental considerations}

\medskip
The $J_z = 0, \gamma\gamma \rightarrow q\overline{q}g$ cross
section, even for small values of $y_{cut}$, is a few per cent of
the $J_z = \pm 2, \gamma\gamma \rightarrow q\overline{q}$ cross
section.  This cross section for bottom and charm quarks, in the
massless approximation is illustrated in Fig.\ 2 along with the
non-radiative backgrounds.  In a photon linear collider, it is
possible to achieve a $\frac{J_z = 0}{J_z = \pm 2}$ ratio of
20(50), so in order to bring the {\it rates} for the radiative
processes down it is necessary to find cuts which further reduce
the radiative backgrounds by a factor of about 5-10, without
seriously degrading the $H \rightarrow b\overline{b}$ signal.  To
explore this in Ref.\ \cite{BSO} a Monte-Carlo integration was
employed (via JETSET 6.3 \cite{TS}) with a simple detector
simulation.

Assuming $W_{\gamma\gamma} \simeq$ 100 GeV and imposing a
$y_{cut}$ of 0.02 with $|\cos \theta| < 0.7$ we found
\begin{eqnarray}
\sigma_{J_z = 0} (\gamma\gamma \rightarrow H \rightarrow
b\overline{b} \rightarrow 2 \hbox{jets}) & = & 0.86 \: pb,
\nonumber \\
\sigma_{J_z = \pm 2} (\gamma\gamma \rightarrow b\overline{b}
\rightarrow 2 \hbox{jets}) & = & 2.21 \: pb, \nonumber \\
\sigma_{J_z = \pm 2} (\gamma\gamma \rightarrow c\overline{c}
\rightarrow 2 \hbox{jets}) & = & 35.6 \: pb, \label{a6} \\
\sigma_{J_z = 0} (\gamma\gamma \rightarrow b\overline{b}g
\rightarrow 2 \hbox{jets}) & = & 0.035 \: pb, \nonumber \\
\sigma_{J_z = 0} (\gamma\gamma \rightarrow c\overline{c}g
\rightarrow 2 \hbox{jets}) & = & 0.87 \: pb. \nonumber
\end{eqnarray}

Although a $y_{cut}$ tends to select very two-jet events the
$q\overline{q}g$ backgrounds (especially in the case of the
Compton-regime configuration) still correspond to a somewhat
different profile of jets.  In addition vertexing proves to be a
very useful tool in separating $b$'s from $c$'s.  In order to
reduce the radiative backgrounds to a manageable level it was
proposed in \cite{BSO} to impose additional event shape, jet
width and vertex cuts.

Applying a sphericity cut of 0.02, a jet width cut of $20^o$,
and requiring 5 tracks with high 3-D impact parameter results in
the following cross sections :
\begin{eqnarray}
\sigma_{J_z = 0} (\gamma\gamma \rightarrow H \rightarrow
b\overline{b} \rightarrow 2 \hbox{jets}) & = & 0.48 \: pb,
\nonumber \\
\sigma_{J_z = \pm 2} (\gamma\gamma \rightarrow b\overline{b}
\rightarrow 2 \hbox{jets}) & = & 1.2 \: pb, \nonumber \\
\sigma_{J_z = \pm 2} (\gamma\gamma \rightarrow c\overline{c}
\rightarrow 2 \hbox{jets}) & = & 0.54 \: pb, \label{a7} \\
\sigma_{J_z = 0} (\gamma\gamma \rightarrow b\overline{b}g
\rightarrow 2 \hbox{jets}) & = & 0.0031 \: pb, \nonumber \\
\sigma_{J_z = 0} (\gamma\gamma \rightarrow c\overline{c}g
\rightarrow 2 \hbox{jets}) & = & 0.0059 \: pb. \nonumber
\end{eqnarray}

Thus, we anticipate that a modern vertex detector should be
capable of achieving the necessary rejection of background events
while remaining reasonably efficient for the signal $b$-quark
events.

This discussion should not be closed without underlying that the
results presented in Refs.\ \cite{BSO,JT} can be considered only
as the starting points for more refined and detailed
investigations.

\newpage
\noindent {\Large \bf Acknowledgements}

\medskip
I wish to thank James Stirling, Jim Ohnemus and especially Doug
Borden for fruitful collaboration.  This work was supported by
the UK Particle Physics and Astronomy Research Council.

\newpage
\noindent {\Large \bf Figure Captions}

\begin{itemize}
\item[Fig.\ 1] Born cross sections for $\gamma\gamma \rightarrow
b\overline{b}$ and $\gamma\gamma \rightarrow c\overline{c}$ in
polarized collisions.  A cut of $|\cos \theta| < 0.7$ has been
applied.  For comparison the Higgs boson signal has been
superimposed with the normalization as described in the text
$\left (\frac{d L}{d W} = 0.2 fb^{-1}/{\rm GeV} \right )$.

\item[Fig.\ 2] Born cross sections for $\gamma\gamma \rightarrow
b\overline{b}$ and $\gamma\gamma \rightarrow c\overline{c}$ and
the radiative background computed in the massless approximation.
\end{itemize}
\end{document}